\documentclass[%
 reprint,
 amsmath,amssymb,
 aps,
]{revtex4-2}

\usepackage{graphicx}
\usepackage{dcolumn}
\usepackage{bm}

\begin{document}

\title{Observation of long-range carrier diffusion in InGaN quantum wells,\\ and implications from fundamentals to devices}

\author{Aurelien David}
 \email{aureliendavid@google.com}
\affiliation{%
 Google, 1600 Amphitheater Pkwy, Mountain View CA 94043, USA
}%

\date{\today}

\begin{abstract}
Photoluminescence measurements on high-quality InGaN quantum wells reveal that carriers diffuse laterally to long distances at room temperature, up to tens of microns. This behavior, which shows a pronounced dependence on the excitation density, contrasts with the common expectation of a short diffusion length. The data is well explained by a diffusion model taking into account the full carrier recombination dynamics, obtained from time-resolved measurements. These observations have important implications for understanding the high efficiency of III-nitride emitters, but also to properly interpret photoluminescence experiments and to design efficient small-scale devices.
\end{abstract}

\maketitle

Carrier transport in InGaN compounds has attracted interest due to the complex physics at play. The energy fluctuations stemming from the distribution of In and Ga atoms can localize carriers on the scale of a few nm. This influences the macroscopic transport properties, including lateral transport in the plane of thin InGaN layers, leading to a non-trivial diffusion coefficient which various groups have studied. Some key experimental techniques are near-field microscopy,\cite{Mensi18} time-of flight measurements\cite{Danhof11,Danhof12} and transient grating spectroscopy\cite{Vertikov99,Aleksiejunas03}. Notably, these studies often assume a fast recombination lifetime --a few ns-- in their analysis, translating into a diffusion length ($L_d$) of a few hundred nm. This may appear reasonable for LEDs in operating conditions. However, as this Article will show, this overlooks the intricacies of InGaN carrier dynamics.

In this Article, we directly observe long-range lateral carrier diffusion in InGaN quantum wells (QW) at room temperature, through simple photoluminescence (PL) experiments. The diffusion length is power-dependent, reaching tens of microns at low power. This data is very well explained when carrier dynamics are considered. We discuss the fundamental and applied consequences of such long-scale diffusion.

\section{Observation of long-range diffusion}

The studied samples, provided by Soraa Inc., are single QW InGaN structures embedded in $p-i-n$ junctions specially designed for PL studies. They are grown by metal-organic chemical vapor deposition on bulk GaN substrates. They have high internal quantum efficiency (IQE). The recombination physics of similar samples were extensively reported in our previous work.\cite{David17a,David19a,David19b,David19_review} The samples are covered with anti-reflective SiO2 coatings to avoid reflection of the exciting laser and emitted light.

\begin{figure*}
\includegraphics[width=\textwidth]{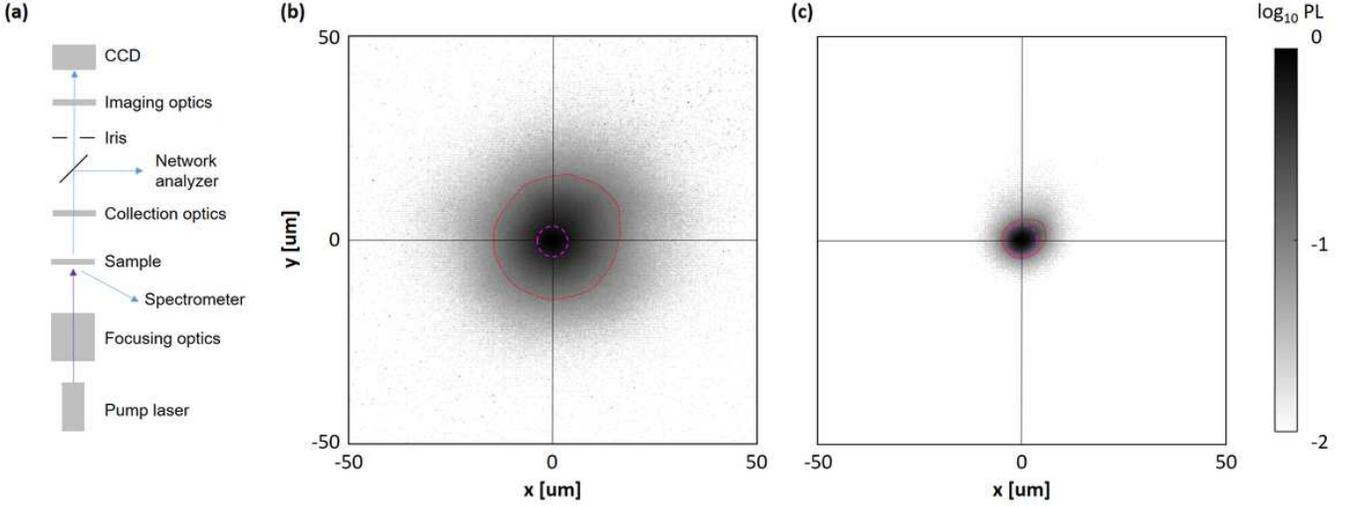}
\caption{\label{fig1} Photoluminescence experiments. (a) Experimental setup. (b-c) CCD image of the PL spot (logarithmic scale, shown on the right) at respective current densities of 25 and 9500 $A/cm^{2}$, respectively. Dotted line: isoline at 10\% of PL intensity. Dashed line: isoline at 10\% of laser excitation intensity.}
\end{figure*}

\begin{figure}
\includegraphics[width=8cm]{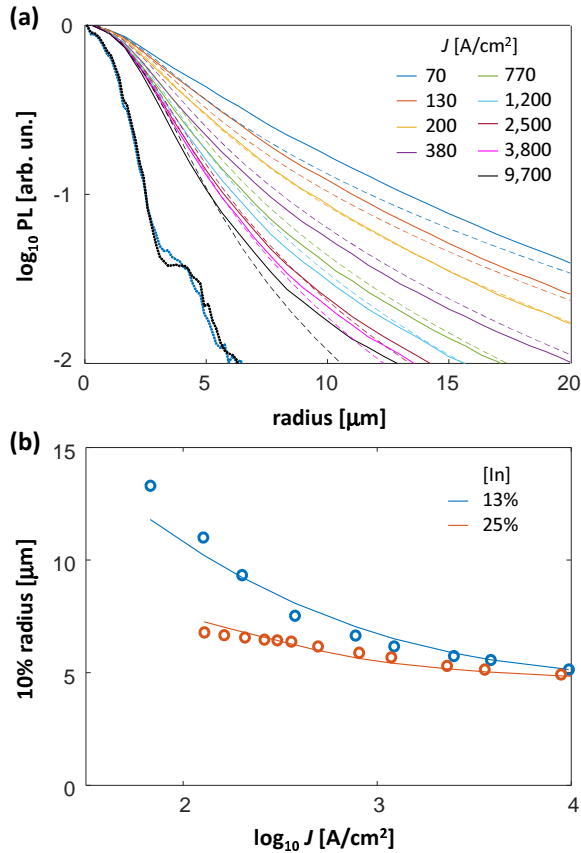}
\caption{\label{fig2} PL diffusion data. (a) Cross-section of PL versus radius, for various current densities $J$. Solid lines: experimental data. Dashed lines: model. Dotted lines: laser profile at the highest and lowest excitation levels. Notice the logarithmic scale. (b) Luminescence spot radius versus current density, for two samples. Circles: experimental data. Lines: model.}
\end{figure}

The PL setup is shown in Fig.~\ref{fig1}(a). The beam of a 405 nm laser diode is expanded and focused to a Gaussian spot with a waist of 2.5 $\mu$m. The laser's drive current is varied to control the excitation density (this does not affect the laser spot's shape, as shown in Fig.~\ref{fig2}(a)). The resulting PL signal is collected by a calibrated spectrometer, which gives access to the absolute PL IQE. Due to the laser's Gaussian spot profile, the excitation density varies spatially; for simplicity, we define a \textit{nominal} photoexcited current density ($J$) in a uniform excitation disc whose diameter is the beam waist. The laser signal can further be time-modulated, and the resulting time-dependent PL collected by a network analyzer to determine the differential carrier lifetime, as described in Ref.~\cite{David17a}. 

A charge-coupled device (CCD) camera is placed in-line with the optical axis to image the laser or PL spot. Laser/PL imaging is selected by inserting appropriate dichroic filters; the image is carefully refocused in each case. 
A small aperture (diameter $\sim$ 2 mm) is placed in the collimated path of the image, to limit the angular aperture of the collected signal to near-normal incidence.


Figs.~\ref{fig1} (b-c) show images of PL spots collected on a sample (4nm-thick QW, composition [In]=$13\%$, peak IQE 80\%) at two current densities. At moderate density (near peak IQE), the spot extends far away from the laser excitation: the radius where the PL reaches 10\% of its peak value (the so-called 10\%-radius) is more than 10 $\mu$m. For comparison, the 10\%-radius of the laser spot, also shown, is about 3$\mu$m. Note that, since luminescence scales with the square of the excitation at low density (in the Shockley-Read-Hall, SRH regime), the PL spot should actually be \textit{narrower} than the laser spot in the absence of diffusion. At a high excitation density (nearly 10 kA/cm$^{2}$), the PL spot is substantially smaller.

These observations are more easily quantified on cross-sectional plots, 
shown in Fig.~\ref{fig2}(a) at various excitation densities (spanning from peak IQE to the droop regime). The PL spot shape strongly depends on the pump power, and extends over tens of $\mu$m at low power. Given the bimolecular nature of radiative recombination, the underlying carrier density extends even farther. The pronounced dependence on the excitation power confirms that optical artifacts cannot account for the large spot size at low excitation. More details are given in the Appendix, where we consider such artifacts in detail (including laser reflection, variations in laser spot size, PL re-absorption, multiple-bounce signal, and thermal effects) and show they do not influence our measurements.

This data cannot be explained assuming a constant $L_d$. Instead, a proper interpretation requires consideration of the power-dependent carrier dynamics. 

To this effect, joint measurements of the IQE and differential carrier lifetime are performed on the sample at varying power density,\cite{David17a} and fitted by the common $ABC$ model to yield recombination coefficients $A$ (SRH), $B$ (radiative), $C$ (Auger). The $ABC$ model is only an approximation, as the radiative and Auger coefficients can display a carrier dependence (as discussed in Refs.~\cite{David17a,David19_BC}). Nonetheless, this carrier dependence is modest in the present sample, and a standard $ABC$ model is suitable. The resulting coefficients are: $A=1.2 \times 10^{5}$ s$^{-1}$, $B = 4 \times 10^{-14}$ cm$^3$s$^{-1}$, $C = 1 \times 10^{-34}$ cm$^6$s$^{-1}$. The very low value of $A$ (corresponding to an SRH lifetime $\sim 10$ $\mu$s) is explained by (1) the high quality of the sample and (2) the thick quantum well, whose low electron-hole overlap reduces the SRH coefficient.~\cite{David17b}

Armed with this knowledge, the PL data is fitted by a steady-state diffusion model:

\begin{subequations}
\begin{eqnarray}
G & = & R - D \Delta n, \label{eq1a}
\\
R & = & An+Bn^2+Cn^3, \label{eq1b}
\end{eqnarray}
\end{subequations}

where $R$ and $G$ are the local recombination and generation rates, $D$ the diffusion coefficient, $n$ the local carrier density. In Eq.~(\ref{eq1a}), $R$ is commonly written as $R=n/\tau$, with $\tau$ (the carrier lifetime) assumed constant. In contrast, in our analysis, $R$ depends on the local carrier density -- in other words, we take into account the carrier-density dependence of the carrier lifetime $\tau(n)$. This variation is very large: at low density (in the SRH regime) $\tau = 1/A \sim 10$ $\mu$s, whereas at high density (in the droop regime) $\tau$ reaches a few ns. This results in a substantial change in the local diffusion length $L_d = \sqrt{D \tau}$, from $\sim 50$ $\mu$m at low density (where $\tau=1/A$) to less than 1 $\mu$m at high density. Note that this model ignores various second-order effects, such as density-induced drift, polarization field screening, and any possible density dependence of $D$.

\begin{figure*}
\includegraphics[width=\textwidth]{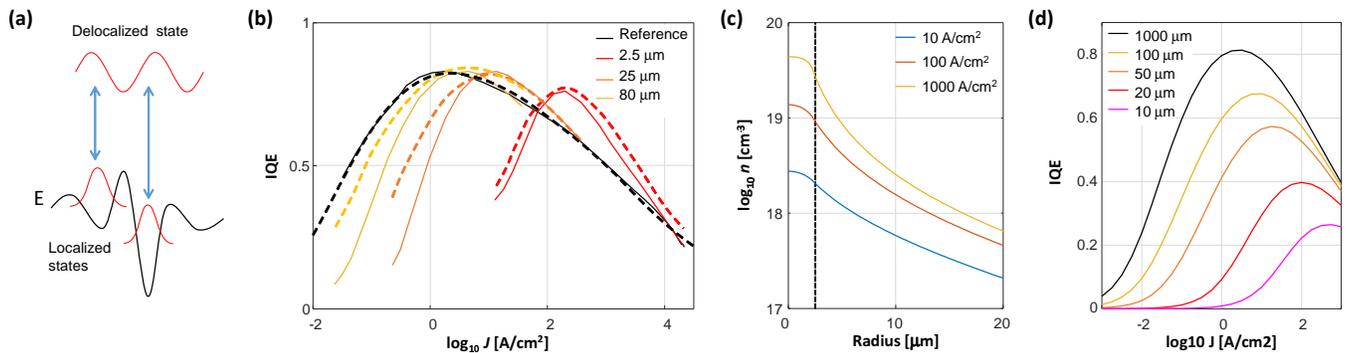}
\caption{\label{fig3} (a) Sketch of thermalization process, driven by fast scattering (arrows) between localized and delocalized states (red shapes). (b) IQE from PL experiments. Solid lines: experimental data. The reference curve is obtained by varying the spot size. The other curves, with a constant spot size and varying power, have an altered shape due to diffusion. Dashed lines: fit by diffusion model. (c) Modeled diffusion of carriers from an injected area with 2.5 $\mu$m radius (injection boundary shown as dashed line) at various current densities. (d) Modeled IQE of LEDs of varying sizes, with non-radiative sidewall recombinations.}
\end{figure*}

The model is solved numerically by finite differences with an iterative loop, solving Eqs.~(\ref{eq1a}) and ~(\ref{eq1b}) in turn until convergence. The spatial distribution of the luminescence is then computed as $Bn^2$. The data to be fitted is the set of PL spot profiles (Fig.~\ref{fig2}) and corresponding IQEs at varying power (shown in Fig.~\ref{fig3}(b)).  This is a very over-specified data set, since the only free parameter is the diffusion coefficient $D$. Good fits are obtained for $D = 3$ cm$^2$s$^{-1}$, a value in good agreement with previous reports\cite{Mensi18,Danhof11,Danhof12,Vertikov99,Aleksiejunas03}. This confirms that the PL spot size is determined by density-dependent carrier diffusion.

Similar measurements are repeated on a sample with a different QW configuration: thickness 2.5 nm, composition [In]$\sim 25 \%$. Despite its high In composition, this sample has a relatively high peak IQE of 73\%. The PL spot size varies with excitation density, albeit in a smaller range than for the previous sample: from 7 $\mu$m at low density to 5 $\mu$m at high density. Again, this data is well explained by a diffusion model based on lifetime measurements. The experimental and modeled 10\%-radius are shown in Fig.~\ref{fig2} (b). This sample is best fitted with a higher diffusion coefficient $D=6$ cm$^2$s$^{-1}$.

Our observations might seem at odds with multiple literature reports of $L_d \sim$ 100~nm.\cite{Rosner97,Vertikov99,Kaneta08} In fact, such a short scale is valid at high excitation density (fast lifetime). In contrast, the present measurements, at larger distances and on a logarithmic scale, reveal the long-range behavior. Besides, samples of sufficient quality (i.e. low $A$ coefficient) are required to observe long-range diffusion -- this was not necessarily the case in past studies.

Neither are our conclusions in conflict with numerous microscopy studies in GaN (e.g. cathodoluminescence) with $\sim$100 nm resolution, because GaN has a much faster SRH lifetime ($\sim$50 ps\cite{Maeda19}) and accordingly a short $L_d$. Regarding microscopy in InGaN, very few studies exist at room temperature in high-IQE material -- both required to observe long-range diffusion. Ref.~\cite{Liu_thesis} did perform such a study and found that sub-micron features become smeared out at $T=170$ K -- in excellent agreement with our findings. Finally, note that imaging resolution around dislocations is not governed by diffusion.\cite{Kaganer19}

In summary, lateral diffusion in InGaN QWs can be substantial. Indeed, even in high carrier density experiments, the lifetime increases rapidly away from the excited region, leading to a large local $L_d$. This effect holds true in high-IQE samples, as it goes hand-in-hand with a slow SRH lifetime, which governs long-scale diffusion. The electron-hole separation, caused by the polarization fields that characterize III-nitride heterostructures, overall slows down the recombination dynamics~\cite{David19_review}: this leads to longer $L_d$ than in conventional III-V samples\cite{Fiore04} and to the pronounced density-dependence reported here. 

\section{Implications of long-range diffusion}

We now discuss three implications of these observations, from fundamental to applied.

\subsection{The IQE of III-nitride quantum wells} 

A long-standing puzzle is how III-nitrides can reach high IQE despite a high threading dislocation density (TDD), as dislocations are believed to act as non-radiative centers. It is often argued that this can be explained by carrier localization at indium-rich fluctuations ($\sim$10 nm scale), which prevent the carriers from reaching dislocations.\cite{Chichibu06} However, the long-range diffusion reported here is incompatible with such tight localization. Accordingly, the Author's recent numerical studies found no support for a fully-localized carrier distribution at room temperature, but rather for a thermalized distribution with a mix of localized and propagative states, as sketched in Fig.~\ref{fig3}(a).\cite{David19b} Incidentally, the value of $D$ for InGaN (both here and in other reports) is on the same order as for bulk GaN\cite{Aleksiejunas03b}, further confirming that alloy disorder does not lead to full localization.

More quantitatively, the present samples were grown on bulk GaN substrates, and have a low TDD $\sim$ 1-5 $\times 10^6$ cm$^{-2}$. This corresponds to a few dislocations in a 10$\times$10 $\mu$m square. It is clear from Fig.~\ref{fig1} that at low-to-moderate excitation density, the carrier distribution will `see' tens of dislocations -- yet, this does not preclude a very high peak IQE. This would be even more true in common GaN-on-sapphire samples, where the TDD reaches $\sim 10^8$ cm$^{-2}$ and high IQE is nonetheless achievable.

It has also been proposed that delocalization at high density could account for IQE droop.\cite{Hammersley12} This is also incompatible with our observations, as diffusion actually decreases at higher density. This causes no contradiction with the previously-reported trend of a higher contrast at higher current in nanoscopic imaging of InGaN luminescence:\cite{Pozina15} this effect is due to a sharpening of the local carrier distribution (due to lower diffusion) which reveals inherent in-plane inhomogeneities.

Therefore, we conclude that localization alone cannot explain the high IQE of III-nitrides. Rather, dislocations must have a limited non-radiative effect -- possibly thanks to the shielding effect of local strain \cite{Liu16} or V-pits \cite{Hangleiter05}. In fact, there is conclusive evidence that SRH recombinations are instead mostly caused by point defects.\cite{Haller17}

\subsection{Interpretation of PL experiments}

The IQE of semiconductors is often studied with power-dependent PL: the excitation is varied to obtain a PL IQE curve without device fabrication. This curve can then be used for temperature-dependent analysis\cite{Watanabe03} or $ABC$ fitting\cite{Karpov15}.

However, our findings imply that the shape of the IQE curve can be strongly affected by diffusion, especially when a focused laser spot size is used (which is desirable to reach the droop regime). This is illustrated in Fig.~\ref{fig3}(b), which shows the PL IQE curve collected with various laser spot sizes. In this comparison, the reference IQE curve (whose details will be explained hereafter) is the PL IQE curve associated with the lifetime measurements; it constitutes a `ground truth' for the other PL measurements. In comparison, the measurement with a maximally-focused laser spot severely alters the IQE curve: the peak IQE is reached at a much higher nominal current density, and the overall shape is much narrower. Qualitatively, it is clear how diffusion causes such distortions: although the laser is tightly focused, the generated carriers diffuse laterally, which results in a much-reduced effective carrier density -- thus the late onset of peak IQE and droop. To verify this quantitatively, we compute the IQE versus excitation density, using Eqs.~(1a-b). The results, shown in Fig.~\ref{fig3}(b), agree very well with the experimental data; they should be compared with the theoretical IQE curve in the absence of diffusion effects, which agrees very well with the reference IQE curve.

As a further confirmation, measurements are repeated with de-focused laser spots under two de-focus conditions, with respective diameters 25 and 80 $\mu$m. The resulting IQE curves are shown in Fig.~\ref{fig3}(b): as the laser is defocused, the IQE curve shape is modified, and progressively converges towards the reference IQE shape. Again, the theoretical IQE curves predicted by the diffusion model are in excellent agreement with measurements. 

Clearly, the distortion effects reported here would substantially alter any analysis of the IQE curve, and should be minimized. This can be done by using a less-focused laser beam (though this may preclude a high excitation density). A better approach is that used to obtain the reference IQE curve, which we will now clarify. This curve is in fact obtained by maintaining a constant (maximal) laser power, while progressively de-focusing the laser spot (in contrast to the conventional measurements where the spot size is constant and the laser power is varied). In this approach, diffusion artifacts are inherently minimized because the excitation spot size increases as the power density decreases: at all but the highest densities, the laser spot is large enough to minimize diffusion artifacts. This advantageous excitation scheme was used by the author in previous studies of carrier dynamics,\cite{David17a,David19a,David19b,David19_review} and is recommended for practitioners of PL measurements.

\subsection{Design of small devices}

There is currently much interest in optical devices with small lateral dimensions, including micro-LEDs and lasers (such as narrow-ridge or VCSELs). In these devices, a small area is injected electrically (e.g. by forming a selective p- contact), with the intention to restrict the QW carrier population to the same area. However, our results indicate that carriers will unavoidably diffuse away from the injected area. Fig.~\ref{fig3}(c) illustrates this. In this calculation, we assume typical recombination coefficients for a thin ($\sim 3$ nm) blue QW (namely  $A=1.2 \times 10^{6}$ s$^{-1}$, $B = 3 \times 10^{-12}$ cm$^3$s$^{-1}$, $C = 1 \times 10^{-31}$ cm$^6$s$^{-1}$, leading to a peak IQE above 80\%).\cite{Galler12,David19_review}  These faster coefficients lead to a shorter $L_d$ than in our first sample -- nonetheless, the low-density $L_d$ remains as high as 17 $\mu$m: this is directly tied to the moderate value of $A$ characteristic of high-IQE material. We assume a cylindrical injection with a 2.5 $\mu$m radius, and compute the radial carrier distribution at various current densities. A substantial diffusion is observed, together with a sharpening of the distribution at high density.

For micro-LEDs, this effect has implications regarding non-radiative recombinations at device sidewalls:  it explains how these can affect the efficiency of relatively-large devices (even hundreds of $\mu$m, Ref.~\cite{Olivier17}) -- which would clearly not occur if $L_d$ were only 100 nm. To confirm this, we model the effective IQE of a blue LED (square shape, same recombination coefficients as above) with sidewall recombinations of infinite velocity, and varying dimensions. The results, shown in Fig.~\ref{fig3}(d), confirm that even for a 100 $\mu$m-wide device, sidewall recombinations significantly affect the IQE. This result is in qualitative agreement with experimental data (e.g. Ref.~\cite{Olivier17}). One could consider adding a non-injected `belt' at the device perimeter to prevent carriers from reaching sidewall defects; even so however, a sizable long-range diffusion current to the non-injected region is unavoidable, and the prospects of this approach are limited. Instead, defect passivation appears more promising.\cite{Wong19}

For laser diodes, the diffusing carriers cause a reduction in injection efficiency near threshold, since they cannot contribute to lasing. The quantitative consequences for threshold density and carrier clamping call for future investigations, especially in emerging small-dimension lasers such as VCSELs.

In summary, long-range diffusion must be accounted for to understand and optimized micron-scale III-Nitride devices.

\section*{Conclusion}

In conclusion, we have observed carrier diffusion over tens of $\mu$m in high-quality InGaN quantum wells under optical excitation, which stands in contrast to the shorter lengths ($\sim 100$ nm) often derived by assuming a nanosecond-order carrier lifetime. Our observations are well explained by a conventional diffusion model, with a reasonable diffusion coefficient, if the full carrier dynamics are taken into account. Indeed, the lifetime varies by orders of magnitude with carrier density, and the diffusion length away from the excited region is governed by the SRH lifetime. Hence, long-range diffusion occurs in high-quality InGaN materials, due to their slow SRH lifetime. This has several important implications: the high IQE of InGaN LEDs cannot be simply ascribed to carrier localization; great care must be taken in avoiding diffusion artifacts in photoluminescence experiments; and upcoming small-scale emitters should be designed with long-range diffusion in mind.

\section*{Appendix - Exclusion of optical artifacts}

To be certain that the CCD images are actually measuring the PL spot, we investigated and excluded possible optical artifacts, as described below.

\subsection*{Laser spot variations}

\begin{figure}[!htbp]
\includegraphics[width=8cm]{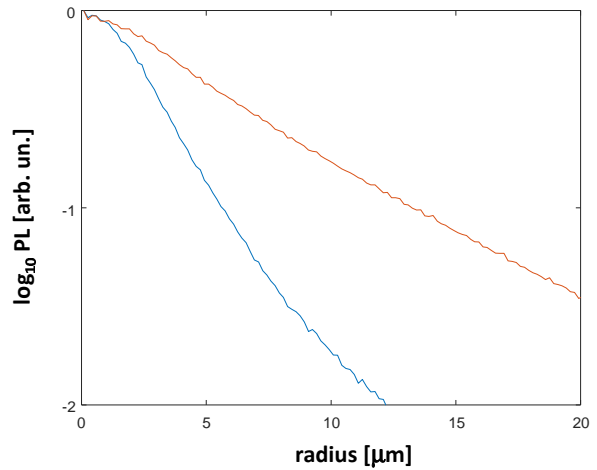}
\caption{\label{SMfig0} Blue: PL profile at maximum power. Red: PL profile with the laser attenuated 300 times by a density filter. The broadening of the PL spot is the same as when the laser power is reduced by controlling the current.}
\end{figure}

One may wonder if the laser spot's profile itself may depend on the laser current. As already shown in Fig.~\ref{fig2}, this is not the case: by imaging the laser spot, we found that it remains stable at all currents. As a further confirmation, we repeated experiments where the laser power, rather than being current-controlled, was decreased by inserting a density filter  in the collimated path of the laser beam (attenuating it 300 times). As shown in Fig.~\ref{SMfig0}, we observed the expected effect: the PL radius increased to 15 $\mu m$ -- the same value as in our manuscript at the same power density.

Although we discuss other possible artifacts below, we believe that the power-dependence of the PL spot, while the laser spot remains the same, is in itself evidence that our observations cannot be caused by an optical artifact (as none would cause such power dependence).

\subsection*{Laser reflection}

The laser beam is partially reflected on both facets of the sample, which may lead to additional excitation with a wider spot size. Hereafter, we show that these reflections do not contribute to our observations.

The sample is coated with SiO$_2$ anti-reflection layers; the layer on the substrate side is optimized for the laser wavelength (to minimize reflection), with a laser reflection of $\sim 1\%$ expected from modeling, and the layer on the epi side is optimized for the PL wavelength (to minimize light extraction effects), with a laser reflection of a few $\%$ expected from modeling.

\begin{figure}[!htbp]
\includegraphics[width=8cm]{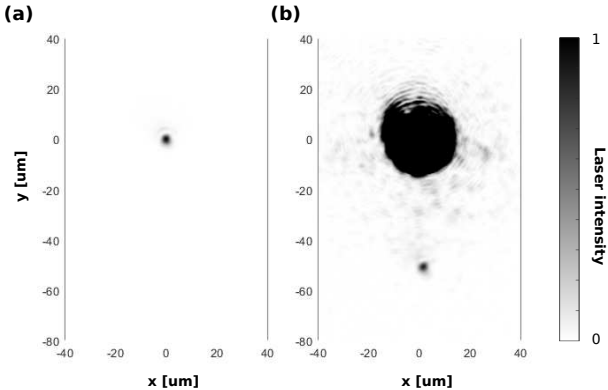}
\caption{\label{SMfig1} Imaging of the laser spot with a tilted sample. (a) Image of the initial laser spot. (b) Image with a much longer integration time. The initial spot is saturated, and the reflection (after bounces off the substrate and epi interfaces) is visible, with an offset due to the sample tilt. The relative intensity of the reflection is 0.3\%.}
\end{figure}

The actual laser reflection can be measured experimentally in our setup. To this effect, the sample is slightly tilted so that the laser spot after a round-trip is offset from the initial laser spot on the CCD image (the offset between the two spots increases with the tilt angle, confirming that the secondary spot is caused by reflections). Note that the focus of the CCD's collecting lens must be slightly adapted to image the laser reflection, due to its extra round-trip in the sample. The intensity of the reflection can be measured by the intensity of the imaged spot; it is about 0.3\% of the laser spot's peak intensity. This is in reasonable agreement with the modeled reflection values (namely, we estimate that the substrate-side reflection is about 3\% and the epi-side reflection is about 10\%). Given the broadening of the laser spot after reflection ($\sim 5$ $\mu$m radius), the relative peak excitation intensity of the reflected spot is well below 1\%, which is too low to induce luminescence on the scale we are observing (our PL data shows two decades of dynamic range).

\begin{figure}[!htbp]
\includegraphics[width=8cm]{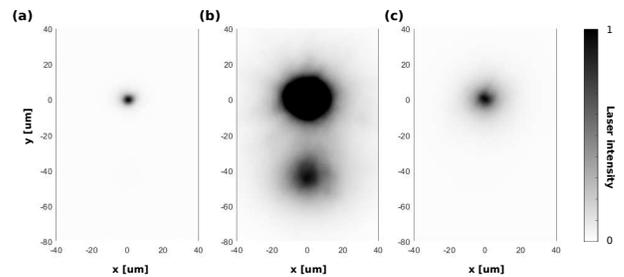}
\caption{\label{SMfig2} PL images in the same tilted-sample geometry as in Fig.~\ref{SMfig1}. (a) At maximum excitation density, the primary PL spot has a radius of $\sim 4-5 \mu$m (same value as for normal incidence). (b) In the same conditions, but with a much longer integration time, the PL from the reflected laser spot can be seen. It has a large radius, due to the low excitation density and the broadening of the reflected laser spot. The relative peak PL intensity is 1\% of the primary PL spot. (c) At lower excitation density, the primary PL spot has a radius of $\sim$15~$\mu$m (same value as for normal incidence).}
\end{figure}

To further confirm unambiguously that this laser reflection does not contribute to the PL measurement, we can actually measure the PL spot size in this tilted-sample geometry. Fig.~\ref{SMfig2} shows such measurements.

Fig.~\ref{SMfig2}(a): at the maximum laser power, the initial PL spot has a 10\%-radius of $\sim 4-5$ $\mu$m. This is the same value as reported in normal-incidence geometry.

Fig.~\ref{SMfig2}(b): at this same power, the PL spot size from the laser reflection (which is of course offset from the initial PL spot) has a large radius, about 15 $\mu$m. This is because the reflected spot is excited at low intensity, and hence is much larger due to diffusion. This figure shows an image with a long CCD integration time, such that the primary PL spot is saturated and the reflected PL spot can be measured.

Fig.~\ref{SMfig2}(c): at the minimum laser power, the initial PL spot has a 10\%-radius of 15 $\mu$m, again the same as for normal incidence.

In summary, if we tilt the sample to move the laser reflection away from the initial laser spot, we obtain the exact same PL spot sizes as reported previously. This confirms laser reflection plays no roles in the results.

\subsection*{PL reabsorption / photon recycling}

\begin{figure}[!htbp]
\includegraphics[width=8cm]{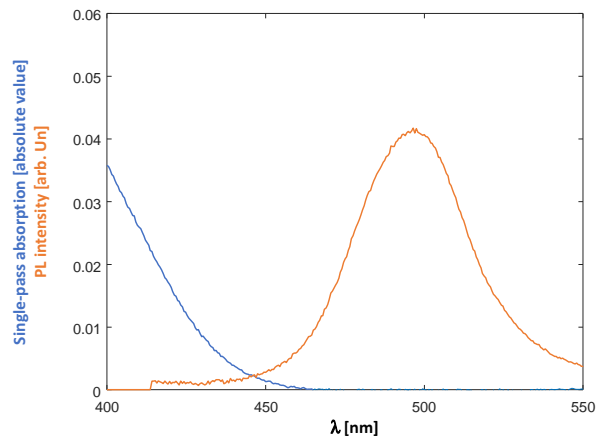}
\caption{\label{SMfig3} PL intensity (at low excitation) and single-pass absorption of the QW. The strong Stokes shift leads to a low net single-pass reabsorption $\sim 0.02\%$.}
\end{figure}

One may wonder if PL guided inside the sample could be re-absorbed by the QW and lead to PL re-emission (a process also known as photon recycling). As we will now show, reabsorption in this sample is very weak and causes no contribution to the PL measurements.

The single-pass reabsorption in this single-QW sample is very low. It can be quantified accurately by measuring the absorption and PL spectra, as shown in Fig.~\ref{SMfig3} (with the PL spectrum shown at low power density).

The corresponding single-pass reabsorption is $\sim$0.02\%. This very low value is due to having only one QW and to the strong Stokes shift in this sample, caused by the p-i-n geometry. The single-pass reabsorption increases at higher power density (because the PL spectrum blue-shifts), and it reaches $\sim$0.5\% at the highest density investigated. 

The resulting net re-excitation of the QW in the experiment is caused by two sources of reflection. (1) PL reflected from the epi surface (located 150~nm away from the QW): this reflection occurs within the same lateral region as the laser spot, because light cannot travel laterally far over such a small thickness. (2) PL reflected from the substrate surface (located $L=150$~$\mu$m away from the QW): this reflection is diluted by the very long distance traveled, and amounts to a very low re-excitation density (scaling with $1/L^2)$.

\begin{figure}[!htbp]
\includegraphics[width=8cm]{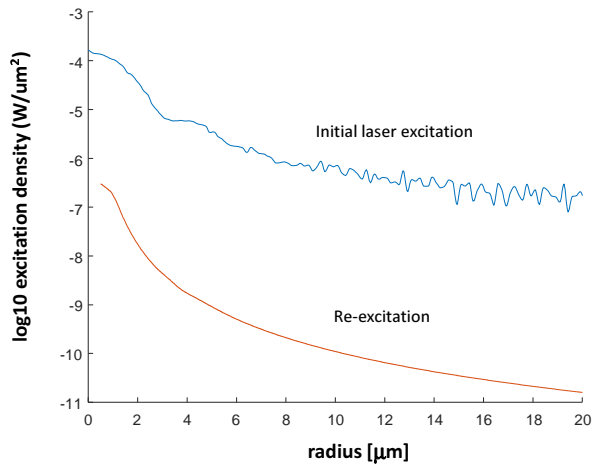}
\caption{\label{SMfig4} Comparison of the experimental initial QW excitation by the laser (blue) and computed the re-excitation due to PL reabsorption (red). Re-excitation is always lower by orders of magnitude, precluding any impact on our results. [Note that the long tail of the initial excitation is caused by background noise of the CCD].}
\end{figure}

These effects are easily quantified by a simple raytracing calculation, taking into account the sample and excitation geometry, and the QW reabsorption. The results are shown in Fig.~\ref{SMfig4}. In this calculation, we use the worst-case single-pass absorption of 0.5\% (corresponding to the highest excitation density). The computed re-excitation is compared with the experimental initial excitation (defined as the product of laser power density times absorption coefficient at the laser wavelength), both in absolute units. Re-absorption is most intense at the beam center (this corresponds to effect (1) above: reflection at the epi surface) and then strongly decreases away from the beam. It is always orders of magnitude smaller than the initial excitation, and therefore does not contribute to the PL signal we measure.

It should be noted that the \textit{total re-excitation} (integrated over the whole sample area) is actually not small -- in this calculation, more than 10\% of the emitted light eventually gets reabsorbed. However, this occurs over the whole area of the sample; therefore, the re-excitation density is very low, as shown above.

\begin{figure}[!htbp]
\includegraphics[width=8cm]{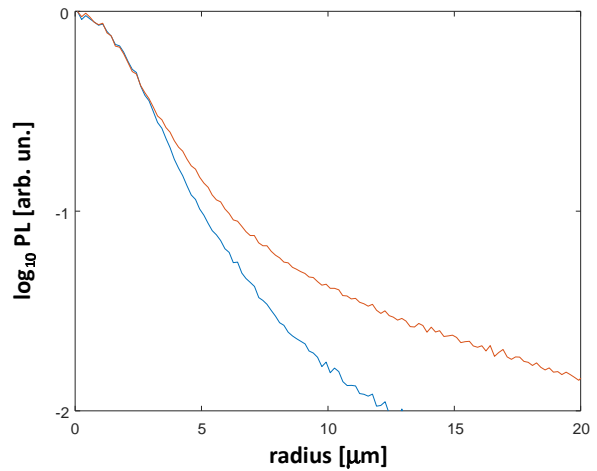}
\caption{\label{SMfig5} PL spot cross-section at the maximum excitation density. Blue: with an iris limiting the numerical aperture of the collection; red: without the iris.}
\end{figure}

\subsection*{Multiple bounces}

Despite the presence of anti-reflection coatings, a fraction of the PL is reflected at interfaces. Therefore, PL emitted off-axis can travel laterally inside the sample, and lead to a broadening / smearing-out of the PL spot. We find that this effect is indeed significant at the highest laser density (where the PL spot is the sharpest), if no precautions are taken. Fig.~\ref{SMfig5} compares two measurements at the highest laser intensity. In one, the numerical aperture of the collecting lens is not limited (leading to a collection $>10^o$), whereas in the other a small iris (diameter 2 mm) placed in the collimated imaging path limits the collection to near normal incidence. The former measurement shows an artificially broad tail caused by multiple-bounce PL. In the latter, off-axis PL is not collected and the intrinsic PL spot shape is resolved. In the rest of this article, an iris was used to remove this artifact.

Note that the iris does not affect the results at lower power density, where diffusion is stronger and the PL spot is broad, so that multiple-bounce PL does not lead to further broadening.

\subsection*{Non-linear absorption}

One might worry that two-photon absorption (2PA) could alter our measurements, leading to a non-linear power dependence. However, 2PA is too low by orders of magnitude to contribute to PL. The 2PA coefficient in InGaN is on the order of $\beta \sim 10$ cm/GW. We are using a 70 mW CW laser diode; even at the peak laser intensity (say with a 1 $\mu$m radius), the resulting 2PA absorption coefficient is thus $\sim 0.02$ cm$^{-1}$ -- to be compared with the linear absorption coefficient $\sim 10^4-10^5$ cm$^{-1}$.

\subsection*{Thermal effects}

\begin{figure}[!htbp]
\includegraphics[width=8cm]{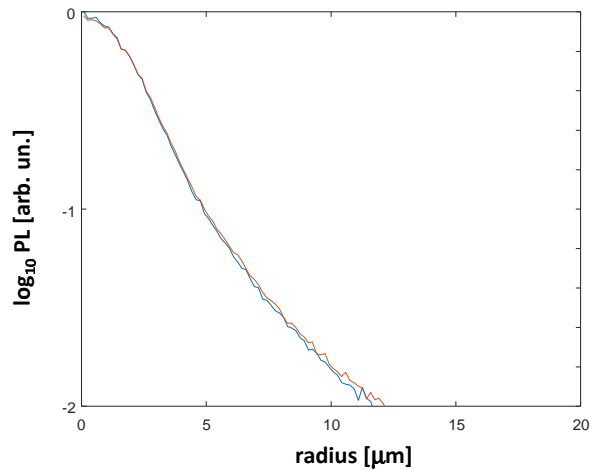}
\caption{\label{SMfig6} PL spot cross-section at the maximum excitation density, with CW and pulsed excitation. The two profiles are identical, showing that no thermal effects influence the results.}
\end{figure}

One might worry that lattice heating at high excitation could influence diffusion. However, the total absorbed laser power is small ($\sim$2.5 mW at the highest density), and we expect it to be dissipated in the GaN substrate with its high thermal conductance. To verify this, we repeated a measurement at the maximum power, comparing the laser in CW mode and pulsed mode (100~ns pulses, 10~$\mu$s period, 1\% duty factor). The PL profiles, shown in Fig.~\ref{SMfig6}, are not affected by the duty factor.
 
%

\end{document}